\documentclass{article}
\begin{document}

\title{Lagrangian description of world-line deviations}

\author{Morteza Mohseni\thanks{email:m-mohseni@pnu.ac.ir}\\\small{Physics department, Payame Noor University, 19395-4697 Tehran, Iran}}

\maketitle
\begin{abstract}
We introduce a Lagrangian which can be varied to give both the
equation of motion and world-line deviations of spinning particles
simultaneously.
%\keywords{World-line deviation \and Spinning particles \and Combined actions}
%\PACS{04.20.-q \and 04.25.g \and 04.30.Nk}
\end{abstract}

\section{Introduction}
The equation of geodesic deviation is a well-known equation of
general relativity with important applications, namely, it
describes the relative motion of many particles. This equation
could be derived by a variety of methods including the second
covariant variation of the point particle Lagrangian. It could
also be derived by first variation of a combined action that gives
both the equation of motion and the deviation equation at once
\cite{baz}. The Lagrangian approach has several advantages, namely
it allows the deviation to be formulated for other dynamical
systems with more general objects \cite{rob}.

In a recent publication \cite{moh} the geodesic deviation equation
was generalized to a deviation equation for the world-lines of
spinning particles. This was basically achieved by considering a
one-parameter family of world-lines of spinning particles and
demanding the invariance of the equations of motion, the
Mathisson-Papapetrou-Dixon equations \cite{dix}, on different
values of the parameter. The same technique was applied in
\cite{hey} to the Dixon-Souriau equations \cite{sou} to obtain the
world-line deviations of charged spinning particles.

In the present work we introduce a combined Lagrangian and show
that both the equation of motion and the deviation equation of
spinning particles can be derived from this Lagrangian. In this
Lagrangian approach the spin is represented by the gyration of an
orthonormal tetrad attached to the particle world-lines.

\section{The action}
We begin with a matter Lagrangian of the form
\begin{equation}\label{e01}
{\mathcal L}=n\sqrt{-g} L(e^a_\mu,{\dot e}_\mu^a,u_\mu)
\end{equation}
where $n$ is the particle number density, $u^\mu$ are the particles
four-velocity satisfying
\begin{equation}\label{e02}
u_\mu u^\mu=-1,
\end{equation}
$\nabla_\alpha e_\mu^a$ are the covariant derivatives and
\begin{equation}\label{e03}
{\dot e}_\mu^a=u^\alpha\nabla_\alpha e_\mu^a.
\end{equation}
We adopt an orthonormal tetrad field given by
\begin{eqnarray}\label{e04}
e_\mu^ae_{a\nu}&=&g_{\mu\nu},\nonumber\\e_{a\mu}e_b^\mu&=&\eta_{ab}
\end{eqnarray}
The action
\begin{equation}\label{e05}
S=\int{\mathcal L}d^4x=\int nL(e^a_\mu,{\dot
e}_\mu^a,u_\mu)\sqrt{-g}d^4x
\end{equation}
describes the dynamics of a dust of particles with spin in a fixed
background space-time. We need not to specify the explicit form of
the Lagrangian here. Taking the relation (\ref{e02}) into account,
we impose the following constraint on the Lagrangian \cite{bai}
\begin{equation}\label{e06}
u^\mu\frac{\partial L}{\partial u^\mu}=0.
\end{equation}
The momentum four-vector and the spin tensor are defined by
\begin{eqnarray}
P^\mu&=&\frac{\partial L}{\partial u_\mu}-Lu^\mu
\label{e17}\\S^{\mu\nu}&=&\frac{\partial L}{\partial{\dot
e}_a^\nu}\,e_a^\mu-\frac{\partial L}{\partial{\dot
e}_a^\mu}\,e_a^\nu\label{e19},
\end{eqnarray}
respectively. Translational equations of motion may be obtained by
extremizing ing the action on variation of world-lines \cite{bai,isr}. For
a one-parameter family of tetrad fields $e^a_\alpha(x,\lambda)$
and congruences $x^\mu(t,\lambda)$ we hold the tetrad attached to
world-lines fixed by parallel propagation. Under infinitesimal
variation
\begin{equation}\label{e30}
x^\mu(t,\lambda)=x^\mu(t,0)+\lambda n^\mu(t),
\end{equation}
various quantities vary as follows
\begin{eqnarray}
\delta(e_\alpha^a)&=&0,\label{e09}\\
\delta(d\tau)&=&-\lambda u^\alpha\nabla_\alpha n_\mu u^\mu d\tau,\label{e10}\\
\delta(u^\mu)&=&\lambda h_\alpha^\mu u^\nu\nabla_\nu n^\alpha,\label{e11}\\
\delta({\dot
e}_\mu^a)&=&\lambda\left({R^\nu}_{\mu\alpha\beta}u^\alpha n^\beta
e_\nu^a+{\dot e}_\mu^au_\nu u^\beta\nabla_\beta
n^\nu\right),\label{e13}\\
\delta(n\sqrt{-g}d^4x)&=&-\lambda u^\mu u_\alpha\nabla_\mu
n^\alpha n\sqrt{-g}d^4x
\end{eqnarray}
in which $h_\nu^\mu=\delta_\nu^\mu+u_\nu u^\mu$ is a projection. The
last equation above follows from the conservation of the number of
particles in an infinitesimal flux tube. Equation (\ref{e13}) may be derived as follows
\begin{eqnarray*}
\delta({\dot e}_\alpha^a)&=&\delta(u^\mu\nabla_\mu
e_\alpha^a)\\&=&\delta(u^\mu)\nabla_\mu
e_\alpha^a+u^\mu\delta(\nabla_\mu
e_\alpha^a)\\&=&\lambda(\delta_\alpha^\mu+u_\alpha u^\mu)
u^\nu\nabla_\nu n^\alpha\nabla_\mu e_\alpha^a+\lambda u^\mu
n^\kappa\nabla_\kappa\nabla_\mu
e_\alpha^a\\&=&\lambda(u^\kappa\nabla_\kappa n^\mu\nabla_\mu
e_\alpha^a+u^\mu n^\kappa\nabla_\kappa\nabla_\mu e_\alpha^a
+u^\kappa\nabla_\kappa n^\beta u_\beta u^\mu\nabla_\mu
e_\alpha^a)\\&=&\lambda(-u^\kappa n^\mu\nabla_\kappa\nabla_\mu
e_\alpha^\mu+u^\mu n^\kappa\nabla_\kappa\nabla_\mu e_\alpha^\mu)+
\lambda{\dot e}_\alpha^au^\beta\nabla_\beta
n_\mu u^\mu+(div.)\\&=&\lambda{R^\beta}_{\alpha\mu\nu}u^\mu n^\nu
e_\beta^a+\lambda{\dot e}_\alpha^au^\beta\nabla_\beta n_\mu u^\mu.
\end{eqnarray*}
With the aid of these relations we are able to calculate the variation of the action. The result is
\begin{equation}\label{e15}
\delta{S}=\lambda\int n\left(\frac{\partial L}{\partial
u^\mu}u^\alpha\nabla_\alpha n^\mu+\frac{1}{2}\left(\frac{\partial
L}{\partial{\dot
e}_{[\mu}^a}e_{\gamma]}^a\right){R^\gamma}_{\mu\nu\alpha}u^\nu
n^\alpha\right)\sqrt{-g}d^4x
\end{equation}
in which use has been made of relation (\ref{e06}). In terms of
momentum and spin tensor, this may be rewritten as
\begin{equation}\label{e20}
\delta{S}=\lambda\int n\left(P_\mu{\dot n}^\mu
-\frac{1}{2}S^{\alpha\beta}{R^\mu}_{\nu\alpha\beta}u^\nu n^\mu\right)\sqrt{-g}d^4x
\end{equation}
or
\begin{equation}\label{e21}
\delta{S}=-\lambda\int nn_\mu\left({\dot P}^\mu
+\frac{1}{2}S^{\alpha\beta}{R^\mu}_{\nu\alpha\beta}u^\nu\right)\sqrt{-g}d^4x+(div.)
\end{equation}
Inspired by this relation, and noting the relation
$n\sqrt{-g}d^4x=Nd\tau$ in which $N$ is the conserved number of
particles in an infinitesimal flux tube, we consider the following
combined Lagrangian
\begin{equation}\label{e23}
S_c=-\int n_\mu\left({\dot P}^\mu
+\frac{1}{2}S^{\alpha\beta}{R^\mu}_{\nu\alpha\beta}u^\nu\right)d\tau.
\end{equation}
This is consistent with the general form of the relative motion Lagrangian
given in \cite{mark}.
\section{Equations of motion and deviation}
The translational equation of motion is obtained from (\ref{e23})
by variation with respect to $n^\mu$. This result in
\begin{equation}\label{e24}
{\dot P}^\mu=-\frac{1}{2}{R^\mu}_{\nu\alpha\beta}u^\nu S^{\alpha\beta},
\end{equation}
the MPD equation of motion for the momentum. A variation of
world-line, of the form of (\ref{e30}) however, results in
\begin{equation}\label{e25}
\frac{D}{D\lambda}\left({\dot P
}^\mu+\frac{1}{2}{R^\mu}_{\nu\alpha\beta}u^\nu
S^{\alpha\beta}\right)=0
\end{equation}
On the other hand, it can be shown that
\begin{equation}\label{e26}
\frac{D}{D\lambda}\frac{D}{D\tau}P^\mu=\frac{D}{D\tau}\frac{D}{D\lambda}P^\mu+{R^\mu}_
{\nu\alpha\beta}u^\nu n^\alpha P^\beta
\end{equation}
where $\frac{D}{D\tau}=u^\alpha\nabla_\alpha,
\frac{D}{D\lambda}=n^\alpha\nabla_\alpha$. By inserting this into
the previous equation and using the abbreviation
$j^\mu=\frac{D}{D\lambda}P^\mu$ we obtain
\begin{equation}\label{e27}
\frac{Dj^\mu}{D\tau}={R^\mu}_ {\nu\alpha\beta}u^\nu n^\alpha
P^\beta+n^\kappa\nabla_\kappa\left(\frac{1}{2}{R^\mu}_{\nu\alpha\beta}u^\nu
S^{\alpha\beta}\right)
\end{equation}
which is the deviation equation, in agreement with the result of
ref. \cite{moh}. It can be shown that the spin equation of motion
\begin{eqnarray}\label{e381}
{\dot S}^{\mu\nu}=P^\mu u^\nu-P^\nu u^\mu
\end{eqnarray}
follows from the fact that the energy-momentum tensor derived from
the action (\ref{e05}) should be symmetric \cite{isr}. One may
incorporate this into the above Lagrangian as a constraint to
obtain the relevant deviation equation.
\section{Discussion}
We started with a Lagrangian whose specific form was not given. We
then showed that its variation with respect to the particle
world-lines lead to a new combined Lagrangian from which both the
equation of motion and the deviation equation can be recovered
simultaneously. This approach benefits from the usual advantages
of Lagrangian formulations, say, it is more powerful for studying
the symmetries of the system. It also allows the deviation
equation to be obtained for more general dynamical systems.

The above combined Lagrangian is re-parametrization invariant.
This allows to switch from the normalized four-velocity to a
non-normalized one which is more convenient when we deal with
spinning particles.

Most of the above procedure could be easily generalized to the
world-line deviations of a Weyssenhoff-type spinning fluid by
starting with a more general Lagrangian which is not necessarily
homogenous in particle number density, and to charged spinning
particles.

\textbf{acknowledgements}

\noindent I would like to thank the Research Council of
Payame Noor University for partial financial support.

\end{document}